\documentclass[journal]{IEEEtran}
\usepackage{amsmath,amssymb,amsfonts}
\usepackage{graphicx}
\usepackage{booktabs}
\usepackage{array}
\usepackage[hidelinks]{hyperref}
\usepackage{xcolor}
\usepackage{textcomp}
\providecommand{\permil}{\text{\textperthousand}}
\usepackage{tikz}
\usetikzlibrary{shapes.geometric, arrows.meta, positioning}
\tikzset{
  fbox/.style ={rectangle, draw, rounded corners=1pt, align=center,
                font=\scriptsize, inner sep=3pt, text width=6.4cm, minimum height=5.5mm},
  fterm/.style={fbox, fill=black!7},
  fdec/.style ={diamond, draw, aspect=2.4, align=center, font=\scriptsize,
                inner sep=1pt, text width=2.3cm},
  far/.style  ={-{Latex[length=1.7mm]}}
}

\begin{document}

\title{Real-Time Location-Aware Demand-Shaping for Power-Constrained AC Railway Corridors}

\author{Márton László Ambrus, Xiao Liu, Stuart Hillmansen, and Zhongbei Tian
\thanks{This work was supported by RSSB (project T1366). \emph{(Corresponding author: M. L. Ambrus)}}
\thanks{The authors are with the Birmingham Centre for Railway Research and Education, University of Birmingham, Birmingham, U.K. (e-mail: m.l.ambrus@bham.ac.uk)}}

\maketitle

\begin{abstract}
Power-constrained 25\,kV AC railway sections, particularly under degraded feeding, are protected today by blunt, section-wide power limits that
penalise every train irrespective of whether it contributes to the binding
condition. This paper presents a real-time, location-aware controller that
restores the electrical feasibility of a feeding section with minimal impact on
the timetable: it curtails only the trains that bind, where and when they bind,
evaluating feasibility and per-train available power online with a solver-free
estimate as an in-loop surrogate for the full power flow. Because the estimate is
accurate on average but slightly optimistic at the binding instants, the
controller screens with a small voltage margin, and a full multi-conductor
power-flow solver confirms the restored feasibility. The resulting
selective-curtailment policy is delivered through a cloud-to-edge connected
driver advisory system. On a representative GB 25\,kV corridor under outage feeding, solver-selected
to be infeasible uncontrolled yet restorable, the controller is compared against
the uncontrolled case, the incumbent static limit, and an offline
genetic-algorithm optimum, with every feasibility figure solver-validated.
The static limit restores feasibility at a large journey-time cost by throttling
the whole section; the location-aware controller restores the same feasibility at
one thirtieth of that cost by advising a single train, and matches the offline
optimum's solution in about a second and a half against the optimiser's minute.
Aggregate peak demand is unmoved, because the active constraint is local
far-field voltage rather than gross demand. All claims are relative to the
baselines on a representative corridor; a specific-route deployment study is
future work.
\end{abstract}


\section{Introduction}
\IEEEPARstart{E}{lectrification} of the railway is central to transport
decarbonisation, yet on parts of the GB network the electrical supply is the
binding constraint on what traffic can be run on electric power rather than
diesel \cite{dft2021tdp,nr2020tdns}. Where a feeding
section is weak, most acutely when a substation is out and the section is fed
from one end, the catenary voltage falls as trains draw power, and operators
protect the supply by limiting traction power. In current practice this limit is
\emph{blunt}: a fixed reduction applied across the whole constrained section, to
every train, for the duration of its transit, regardless of whether that train
is responsible for the depressed voltage at any instant.

That bluntness is wasteful, and the reason is structural. The minimum section
voltage is governed not by aggregate power but by the sum, over the trains
sharing a segment, of the product of each train's power and its electrical
distance from the supply \cite{paper1_estimator}. The binding condition is
therefore caused by particular trains, at particular locations, during
particular overlaps of their high-power windows. A static, section-wide limit
ignores this entirely, throttling a lightly loaded train near the supply exactly
as much as a heavily loaded train far out, and so spending far more journey time
than the electrical relief it buys. Operational measurements show the signature
of this inefficiency directly: a blunt limit sharply reduces the short-window
(one-minute) RMS demand while barely changing the long-window (thirty-minute)
RMS, and hence the energy drawn \cite{rssb_t1331}. The energy is merely spread in
time, which is what a demand-side controller seeks, but achieved by the most
expensive available means.

This paper asks whether the same supply protection can be obtained at a fraction
of the journey-time cost by limiting only the trains and instants that bind. The
obstacle to doing so online has been computational: identifying the binding
trains requires evaluating the multi-train power flow, and a full solver is too
slow to run inside a control loop over a forecast horizon. A companion
paper \cite{paper1_estimator} removed that obstacle with a solver-free estimate
of the power available to each train on a feeding section, accurate enough and
fast enough to be queried repeatedly online. The present paper builds the
controller that acts on that estimate, and evaluates it.

The contributions are: (i) a control formulation that casts supply protection as
restoring compliant supply voltage ($V_{\min}\ge U_{\min}$) at the smallest
necessary journey-time penalty, with feasibility and available power evaluated
online by the estimate of \cite{paper1_estimator} (Section \ref{sec:control});
(ii) a transparent, executable selective-curtailment policy that screens with a small
voltage margin to remain robust to the estimate's optimism at the binding
instants, and its delivery through a cloud-to-edge C-DAS architecture
(Section \ref{sec:control}); and (iii) a case study on a representative GB
25\,kV corridor under outage feeding, selected with the full power-flow solver in
the loop and with every feasibility figure solver-validated, comparing the
controller against the uncontrolled case, the incumbent static limit, and an
offline GA optimum, with every parameter traceable to a standard or cited source
(Sections \ref{sec:setup}--\ref{sec:results}, with the scoring and validation
procedure in Section \ref{sec:repro}).
Section \ref{sec:discussion} discusses validity and deployment, and
Section \ref{sec:conclusion} concludes.

\section{Background and Related Work}
\label{sec:background}
\textbf{Energy-efficient train control.} A large literature optimises a single
train's speed profile for energy, from the optimal-control foundations through
to practical advisory systems \cite{SCHEEPMAKER2017355,
albrecht2016key,6410425,8649821,liu2025comparative}.
Multi-train extensions coordinate several trains, typically for energy or
regenerative exchange, and jointly optimise speed profile and timetable
\cite{7039269,6856141,9006917,ran2020energy}; broader assessments catalogue the
measures available to reduce network traction energy \cite{douglas2015assessment}.
These methods largely treat the electrical supply as adequate; the
problem here is the opposite: the supply is the constraint, and the objective
is to protect it with minimal impact on the timetable.

\textbf{Power-flow-coupled optimisation.} Where the supply is modelled,
optimisation is generally offline, coupling a timetable optimiser to a railway
power-flow solver \cite{en13112788,9006917}. Offline DP and GA methods can
approach the true optimum with full future knowledge \cite{haahr2017dynamic,
zhao2014bruteforce,bocharnikov2007optimal}, but cannot run within live-traffic
timescales, and their raw schedules are volatile and hard for a driver to
follow. Accelerating AC power flow for online use is an active topic in the wider
power-systems literature, building on the classical optimal-power-flow
formulation \cite{dommel1968optimal,peterson1972iterative,mohamed2017modified,
zamzam2020learning,zhou2020data}; the estimate of \cite{paper1_estimator} is the
railway-specific counterpart exploited here.

\textbf{Driver advisory systems.} Connected DAS deliver online advice to
drivers and are increasingly deployed \cite{cdas_das}. The gap this paper
addresses is that no deployed DAS shapes \emph{multi-train} demand using a
\emph{location-aware} estimate of available power, in real time, under supply
constraint. A systematic review of algorithmic energy management in constrained
traction networks situates this gap \cite{ambrus2026review}.

\section{System Model and Available-Power Estimate}
\label{sec:model}
The corridor is modelled as a sequence of feeding sections delimited by neutral
sections; within a section the catenary, rails and (where present)
autotransformers form a multi-conductor chain circuit solved by a fixed-point
AC power flow \cite{powerflow_railway,mellitt1978simulator,hu2024traction}. Each train is a complex power injection
at its instantaneous position, subject to the voltage-dependent automatic
current limitation (ACL) of EN\,50388-1: the permissible current derates once
the pantograph voltage falls below $V_1\!\approx\!19$\,kV and is fully
suppressed below $V_0\!\approx\!12.5$\,kV, while $U_{\min}\!=\!17.5$\,kV is the
EN\,50163 compliance limit between them \cite{en50163,en50388}. Train motion
follows the standard distance-domain equations, with tractive effort limited by
the lesser of adhesion and power-over-speed. This solver is both the model to
which the companion estimate \cite{paper1_estimator} is calibrated and the
ground truth against which every committed plan is judged
(Section \ref{sec:setup}).

The physics the controller exploits is captured by one relation. For a train at
distance $d$ from the supply point drawing active power $P$ at pantograph
voltage $V$, on a feeder of series resistance $r$ per unit length, the resistive
voltage drop is approximately
\begin{equation}
V_s - V \;\approx\; r\,d\,\frac{P}{V},
\label{eq:drop}
\end{equation}
so the depression scales with the \emph{product} of power and distance, and when
several trains share the feeder their drops superpose over the shared
path \cite{paper1_estimator}. Which train binds therefore depends on where each
train is, not merely on how much it draws, the structural fact a section-wide
limit ignores.

\subsection{Solver-free available-power estimate}
The control loop of Section \ref{sec:control} does not call the solver online.
Instead it uses the solver-free estimate of \cite{paper1_estimator}, whose
operative form is reproduced here so that the present paper is self-contained.
For $N$ trains at distances $d_i$ drawing complex powers $S_k=P_k(1+\mathrm{j}\kappa)$,
with $\kappa$ fixed by the displacement factor, the train voltages satisfy the
shared-path relation
\begin{align}
V_i &= V_s - \sum_{k} M_{ik}\!\left(\frac{S_k}{V_k}\right)^{\!*}\!,\notag\\
M_{ik} &= \gamma(d_i,d_k)\,Z\big(\!\min(d_i,d_k)\big),
\label{eq:sharedpath}
\end{align}
in which $Z(d)=Z_0+\beta(d)\,z\,d$ is the calibrated series impedance from the
supply point to distance $d$, and
$\gamma(d_i,d_k)=\gamma_\infty+(1-\gamma_\infty)\,e^{-|d_i-d_k|/L}\le 1$
(with $\gamma\equiv1$ on the diagonal) is a separation-dependent mutual
coupling-reduction factor. The generalisation of \eqref{eq:drop} that
invalidates any additive headroom rule is thus retained: train $k$'s current
depresses train $i$'s voltage through the feeder length the two share back to
the supply point. Both terms are calibrated offline, per feeding configuration:
a single-train solver sweep fixes the self-impedance profile $\beta(d)$, and a
small set of two-train sweeps fits $(\gamma_\infty,L)$; for the single-end-fed
corridor used here, $\gamma_\infty\!\approx\!0.02$ and $L\!\approx\!3.4$\,km.
Online, \eqref{eq:sharedpath} is solved by a damped fixed point in a few
$\mathcal{O}(N^2)$ iterations, returning the section minimum voltage
$V_{\min}=\min_i|V_i|$; a bisection on train $i$'s power, re-solving
\eqref{eq:sharedpath} at each trial, returns the power $H_i$ available to train
$i$ before $V_{\min}$ reaches $U_{\min}$. Both are obtained without a power-flow
solve, at a cost that scales with the number of trains rather than with the
network, low enough to evaluate repeatedly inside a controller.

Against matched multi-train power-flow snapshots the estimate reproduces the
per-train available power to a mean absolute deviation of about 9\% (maximum
16.6\%), as a tight two-sided approximation that runs slightly optimistic for
distant, lightly loaded trains \cite{paper1_estimator}, precisely the
configuration that decides feasibility on a single-end-fed section, which is why
the controller of Section \ref{sec:control} screens with a voltage margin. We
take $V_{\min}$ and $H_i$ from this estimate as the controller's in-loop
evaluator of feasibility and headroom, and reserve the full solver for offline
ground-truth validation of the chosen control.

\section{Demand-Shaping Control}
\label{sec:control}
\subsection{Formulation}
Let $P_i(t)$ be the traction power of train $i$ (negative when regenerating) and
$\mathcal{P}(t)=\sum_i P_i(t)$ the aggregate demand presented to the section.
With the $\tau$-window RMS
\begin{equation}
\overline{\mathcal{P}}_\tau(t)=\sqrt{\tfrac{1}{\tau}\textstyle\int_{t-\tau}^{t}\mathcal{P}(s)^2\,\mathrm{d}s},
\label{eq:rms}
\end{equation}
the short window $\tau_1$ (one minute) tracks the instantaneous stress that
depresses voltage, and the long window $\tau_{30}$ (thirty minutes) tracks the
energy the substation rating must accommodate.

On a power-constrained section under outage feeding, the binding operational
limit is not the aggregate demand but the catenary voltage: normal traffic
drives $V_{\min}$ below $U_{\min}$ in the far field, so the section is
\emph{infeasible}. The controller's objective is therefore to restore
feasibility with minimal impact on the timetable, to choose the per-train
profiles $\{P_i\}$, within the operational freedom the timetable allows, that
\begin{equation}
\min_{\{P_i(\cdot)\}}\ \textstyle\sum_i \Delta T_i
\label{eq:objective}
\end{equation}
subject to, for every train $i$ and time $t$,
\begin{align}
V_{\min}(t)\ \ge\ U_{\min},\quad
P_i(t)\ \le\ H_i(t),\quad
\Delta T_i\ \le\ \Delta T_{\max},
\label{eq:constraints}
\end{align}
together with the train-dynamics relations linking $P_i$ to motion and arrival
time. Here $\Delta T_i$ is the journey-time penalty of train $i$ relative to its
unconstrained run and $\Delta T_{\max}$ the tolerance (tens of seconds). The
feasibility and headroom constraints in \eqref{eq:constraints} are evaluated by
the estimate of \cite{paper1_estimator}, which is what makes the problem solvable
online; and because $H_i$ is location-aware, the controller caps only the binding
trains rather than the whole section.

The peak short-window RMS $\overline{\mathcal{P}}_{\tau_1}$ of \eqref{eq:rms} is
retained as a \emph{secondary} demand-shaping indicator. It responds when the
binding condition coincides with the aggregate-demand peak, but on a section
whose voltage binds far from the feed it can be largely unaffected even as
feasibility is fully restored, because the aggregate peak is a whole-line sum
while the binding is local. We report it, and return to this distinction in
Section \ref{sec:discussion}.

\subsection{Selective curtailment policy}
\begin{figure*}[t]\centering
\begin{tikzpicture}[node distance=5mm]
\node[fterm] (start) {Nominal plan (all trains uncapped); order trains by band-entry time};
\node[fbox, below=of start] (take) {\textbf{Restriction pass:} take the next train $i$ in entry order};
\node[fbox, below=of take] (fc) {Forecast the section minimum voltage over train $i$'s traversal of the band (analytic estimate)};
\node[fdec, below=6mm of fc] (dv) {min $V<U_{\min}$?};
\node[fdec, below=10mm of dv] (more) {trains left?};
\node[fbox, below=8mm of more] (relax) {\textbf{Relaxation pass:} loosen each train's cap one level while its traversal stays feasible; repeat until no cap changes (fixed point)};
\node[fterm, below=of relax] (out) {Issue one advisory (committed cap level) per train};
\draw[far](start)--(take);
\draw[far](take)--(fc);
\draw[far](fc)--(dv);
\draw[far](dv)-- node[right,font=\scriptsize]{\,no (feasible)} (more);
\draw[far](more)-- node[right,font=\scriptsize]{\,no} (relax);
\draw[far](relax)--(out);
\draw[far] (dv.east) -- ++(3.6,0) node[midway,above,font=\scriptsize]{yes: tighten train $i$'s cap one level} |- (fc.east);
\draw[far] (more.west) -- ++(-3.3,0) node[midway,above,font=\scriptsize]{yes} |- (take.west);
\end{tikzpicture}
\caption{Online controller logic. A non-anticipative restriction pass restores
feasibility train-by-train at band entry; a relaxation pass then removes the
over-capping by loosening each cap to the least restriction the rest of the
traffic allows. Both passes query only the analytic estimate, so the plan is
computed in milliseconds.}
\label{fig:ctrl_flow}
\end{figure*}
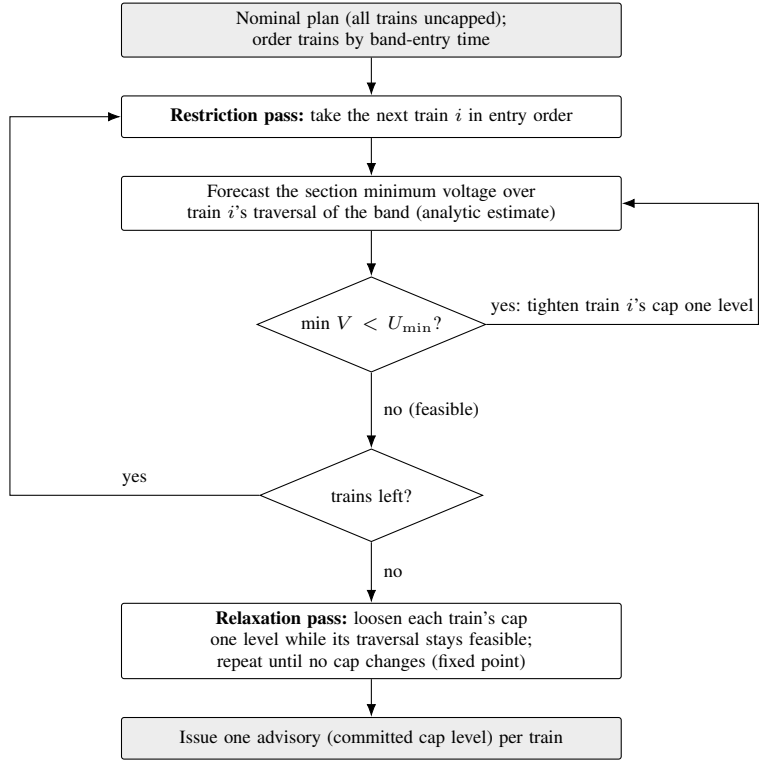

The general solution of \eqref{eq:objective}--\eqref{eq:constraints} is a
constrained optimisation over the joint cap choice. We adopt a transparent,
deterministic policy that exposes the mechanism and is cheap enough to run
online, in two passes (Fig. \ref{fig:ctrl_flow}). The policy is
\emph{non-anticipative}: trains are processed in the order they enter the
constrained band, and a train's cap is decided at its entry, when the capped and
uncapped profiles still coincide and the advice is actionable.

In the \emph{restriction pass}, each train in turn is checked by evaluating the
estimate over its traversal of the band; while the predicted section minimum
voltage is below $U_{\min}$, the train's cap is tightened one level and the
traversal re-checked, until it is feasible or the deepest cap is reached. This
restores feasibility but over-caps, because an early train forecasts against a
still-uncapped future. The \emph{relaxation pass} then loosens each train's cap
to the least restrictive level that keeps its traversal feasible given the
others, and repeats until no cap changes, a fixed point at which every train
carries only the curtailment the others do not already provide. Because
curtailment is concentrated on the binding trains and instants, its journey-time
cost falls where it does electrical work, in contrast to the section-wide penalty
of a static limit. Both passes use only the estimate, so the whole decision costs
milliseconds; the relaxation is a planning-time iteration, so the issued advice
remains a single stable figure per train.

\emph{Screening margin.} The estimate tracks the full power flow closely on
average (Section \ref{sec:model}), but runs optimistic at
the most heavily loaded far-field instants, exactly those that decide
feasibility (Section \ref{sec:results}). A controller that screened at $U_{\min}$
directly would therefore under-cap, leaving a residual breach the solver would
still see. We close this by screening against $U_{\min}+\Delta$ with a fixed
margin $\Delta$ (here $1.5$\,kV): the restriction and relaxation passes both
target $U_{\min}+\Delta$ in the estimate, so that the true minimum voltage clears
$U_{\min}$ once the solver is consulted. The margin is the honest, quantified
price of using a fast estimate near the limit rather than a solver in the loop;
its journey-time cost is small (Section \ref{sec:results}).

\subsection{Executability and C-DAS delivery}
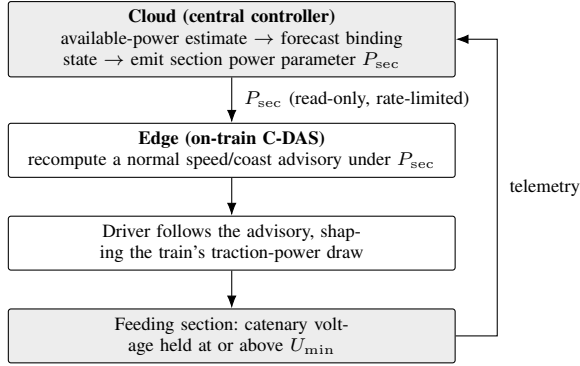
\begin{figure}[t]\centering
\begin{tikzpicture}[node distance=5mm]
\node[fterm, text width=5.7cm] (cloud) {\textbf{Cloud (central controller)}\\ available-power estimate $\rightarrow$ forecast binding state $\rightarrow$ emit section power parameter $P_{\mathrm{sec}}$};
\node[fbox, below=6mm of cloud, text width=5.7cm] (edge) {\textbf{Edge (on-train C-DAS)}\\ recompute a normal speed/coast advisory under $P_{\mathrm{sec}}$};
\node[fbox, below=of edge, text width=5.7cm] (driver) {Driver follows the advisory, shaping the train's traction-power draw};
\node[fterm, below=of driver, text width=5.7cm] (net) {Feeding section: catenary voltage held at or above $U_{\min}$};
\draw[far] (cloud) -- node[right,font=\scriptsize]{\,$P_{\mathrm{sec}}$ (read-only, rate-limited)} (edge);
\draw[far] (edge) -- (driver);
\draw[far] (driver) -- (net);
\draw[far] (net.east) -- ++(0.55,0) |- node[right,font=\scriptsize,pos=0.25]{\,telemetry} (cloud.east);
\end{tikzpicture}
\caption{Cloud-to-edge C-DAS delivery. The central function forecasts the binding
state from the available-power estimate and pushes a single section power
parameter $P_{\mathrm{sec}}$ to each train; the on-train C-DAS turns it into a
normal driving advisory, and train telemetry (position, speed, power) returns to
close the forecast loop. The boundary is one-way and read-only, and the edge holds
its last valid parameter if the link drops.}
\label{fig:cdas}
\end{figure}

Advice that changed every second would be electrically optimal and operationally
useless. The controller therefore commits a power figure over a section and time
window, rate-limited and held with hysteresis, expressed in terms the driver
already uses. It is delivered through the cloud-to-edge C-DAS of
Fig. \ref{fig:cdas}: a central (cloud) function forecasts the binding state with
the estimate and emits a section power parameter; the on-train (edge) C-DAS
recomputes a normal speed/coast advisory under that parameter, which the driver
follows, shaping the train's traction-power draw. Train telemetry (position,
speed, power) returns to the central function to close the forecast loop. The
control boundary is one-way and read-only: the central function sets a parameter,
never a direct command. The device degrades gracefully on loss of connectivity
by holding its last valid parameter, and no new advisory class is introduced. This
is the deployment vehicle, not the contribution.

\section{Case Study Setup}
\label{sec:setup}
\subsection{Representative corridor}
The corridor is a representative GB 25\,kV AC section under degraded (N$-$1)
feeding: 100\,km, direct-fed from one end, with paralleling posts and a
representative line impedance, voltages and current-limitation characteristic
per EN\,50163 and EN\,50388-1. No single real route is reproduced; every parameter is
traceable to a standard or to the cited AC-supply
literature \cite{white2015ac,oura1998railway,bhargava1999railway,powerflow_railway},
so the corridor is a defensible composite. Table \ref{tab:corridor} lists the
parameters and their provenance. The route carries a bidirectional peak-hour
timetable of one high-speed and five regional services at four-minute headway.
The traffic level was chosen \emph{with the full power-flow solver in the loop},
by sweeping headway, mix and count and retaining a case the solver confirms is
genuinely infeasible uncontrolled, its far-field voltage dipping below $U_{\min}$
without collapsing, yet restorable by curtailing the binding train. This matters:
the online estimate is optimistic at the binding instants, so a scenario selected
on the estimate alone can read as restorable while the solver shows it collapsed;
selecting on the solver avoids that trap and makes every feasibility figure below
solver-backed. The single high-speed service is the binder. Its high-power
far-field pass, thickened by regional overlap, drives the dip, and the binding
band sits at 52--76\,km.

\begin{table}[t]
\caption{Representative corridor parameters and provenance.}
\label{tab:corridor}
\centering\small
\begin{tabular}{@{}p{0.30\linewidth}p{0.28\linewidth}p{0.30\linewidth}@{}}
\toprule
Parameter & Value & Source \\
\midrule
Voltages ($V_{oc}/V_n/U_{\min}$) & 27.5 / 25 / 17.5\,kV & EN\,50163 \cite{en50163} \\
ACL onset / cut-off & 19 / 12.5\,kV & EN\,50388-1 \cite{en50388} \\
Power factor & 0.96 & EN\,50388-1 \cite{en50388} \\
Feeding & single-end, DF (outage) & repr. \cite{powerflow_railway} \\
Section length & 100\,km & repr. (matched to route) \\
Line speed & 160\,km/h & typical GB main line \\
Station spacing & $\sim$33\,km & typical GB intercity \\
Gradients & $\le 10\,\permil$ & within GB ruling grades \\
Rolling stock & 1 high-speed, 5 regional & published characteristics \\
Headway / directions & 240\,s / bidirectional & solver-selected (Sec. \ref{sec:setup}) \\
\bottomrule
\end{tabular}
\end{table}

\subsection{Baselines and metrics}
The controller is compared against four references: the \emph{uncontrolled} case
(nominal profiles, no management); the \emph{static section limit}, reproducing
the incumbent blunt reduction (every train capped to $0.85$ of rated power for
its transit of the section); the \emph{offline GA optimum}, a strong reference
computed with full future knowledge \cite{haahr2017dynamic,
zhao2014bruteforce}; and the proposed \emph{online controller}. Metrics follow the
objective: constraint-violation duration, i.e.\ the number of seconds with
$V_{\min}<U_{\min}$ (primary: feasibility restoration is the task); total
journey-time penalty $\sum_i\Delta T_i$ (the cost of restoring it); decision
runtime (the real-time test); and the number of trains advised and cap depth
(advice volatility, the executability measure). Peak one-minute RMS
of \eqref{eq:rms} is reported as a secondary demand-shaping indicator.

The comparison tests two pre-stated hypotheses. $H_{1a}$: the online controller
restores feasibility at least as well as the incumbent static limit, at lower
total journey-time cost. $H_{1b}$: the online controller matches the offline GA
optimum's feasibility and cost, in real time and with few advisories. Both are
decided on solver voltages (Table \ref{tab:hyp}).

\subsection{Validation strategy}
All evaluation is in-simulation; the sim-to-real gap is a declared limitation,
not a hidden assumption. Within that scope, the network model is anchored to an
independent baseline established in prior RSSB work on network-wide
traction-energy management \cite{RSSB_1270} rather than to its own
self-consistency \cite{rssb_t1331}, and the controller is benchmarked against the
\emph{real} incumbent measure rather than a strawman. Crucially, feasibility is
not asserted from the online estimate: every committed plan (uncontrolled,
static, GA and controller) is re-scored second by second by the full
multi-conductor AC power-flow solver, and it is those solver voltages, not the
estimate's, that appear in the results (Section \ref{sec:results}). The estimate
is used only online, to \emph{form} the control; the solver is the judge. To keep
the static and GA baselines like-for-like with the controller, all three screen
against the same $U_{\min}+\Delta$ margin. All claims are relative; absolute MW
figures for any specific real route are out of scope.

\section{Computational Procedure}
\label{sec:repro}

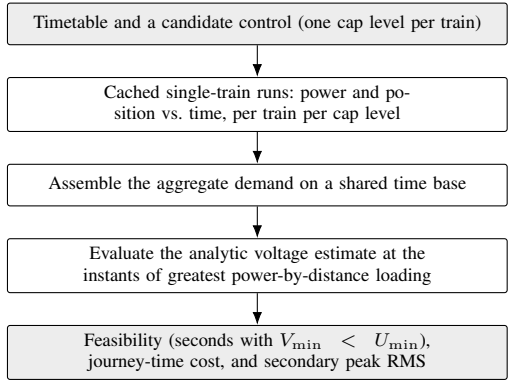
\begin{figure}[t]\centering
\begin{tikzpicture}[node distance=4.2mm]
\node[fterm] (a) {Timetable and a candidate control (one cap level per train)};
\node[fbox, below=of a] (b) {Cached single-train runs: power and position vs.\ time, per train per cap level};
\node[fbox, below=of b] (c) {Assemble the aggregate demand on a shared time base};
\node[fbox, below=of c] (d) {Evaluate the analytic voltage estimate at the instants of greatest power-by-distance loading};
\node[fterm, below=of d] (e) {Feasibility (seconds with $V_{\min}<U_{\min}$), journey-time cost, and secondary peak RMS};
\draw[far](a)--(b); \draw[far](b)--(c); \draw[far](c)--(d); \draw[far](d)--(e);
\end{tikzpicture}
\caption{Scoring a candidate control. Single-train runs are simulated once and
cached, so evaluating a control is a table lookup and one estimate pass, not a
re-simulation, which is what makes the online search tractable.}
\label{fig:pipeline}
\end{figure}

Every candidate control is a choice of one cap level per train from the discrete
set of Section \ref{sec:setup}. It is scored as in Fig. \ref{fig:pipeline}: the
aggregate demand is assembled from the constituent single-train runs on a shared
time base, and the analytic estimate of Section \ref{sec:model} is evaluated at
the instants of greatest power-by-distance loading. Feasibility is the number of
seconds at which the estimated section minimum voltage falls below $U_{\min}$,
and cost is the summed journey-time penalty relative to the unconstrained runs.
The single-train runs are produced once per train per cap level by a
distance-domain traction simulator that applies the section power cap, and
cached, so that scoring a candidate is a lookup and an estimate evaluation rather
than a re-simulation, which is what keeps the controller and the baselines
tractable.
The binding band is located automatically as the stretch of track occupied by
the trains that dominate the lowest-voltage instants. The static limit, the
offline GA and the online controller all draw from this same cached set and the
same estimate, screening at the same $U_{\min}+\Delta$ margin, so their
comparison is like-for-like. Runtime (Table \ref{tab:headline}) is the wall-clock
cost of forming the control decision, the estimate evaluations and ranking, measured
against the GA's total search time on the same scenario.

Two uses of the full multi-conductor solver sit around this fast inner loop.
First, the scenario itself is \emph{selected} with the solver: candidate traffic
levels are swept and each is scored by the solver uncontrolled, retaining one that
is genuinely infeasible yet convergent rather than one the estimate merely reports
as infeasible. Second, every committed plan is \emph{re-scored} by the solver
second by second over the whole horizon, and it is those voltages that populate
Tables \ref{tab:headline} and \ref{tab:solvercheck}. The estimate forms the
control; the solver judges it. The controller, estimator, baselines, solver
harness and representative corridor are available from the authors on request.

\section{Results}
\label{sec:results}
\emph{Every figure in Table \ref{tab:headline} is the full solver's verdict on the
committed plan, over the whole horizon at one-second resolution; the estimate's
value is shown alongside in Table \ref{tab:solvercheck} to quantify its accuracy.
On this corridor there is no voltage collapse under any policy (the scenario was
selected to be infeasible but supplyable), so the infeasible-second count is a
clean measure of the task.}

\begin{table}[t]
\caption{Headline metrics across baselines, \emph{solver-validated}. Primary axes
are feasibility (seconds with $V_{\min}<U_{\min}$) and journey-time cost; minimum
voltage and runtime support them; peak one-minute RMS is a secondary indicator.
The static, GA and controller plans all screen at the same $U_{\min}+\Delta$
margin ($\Delta=1.5$\,kV).}
\label{tab:headline}
\centering\small\setlength\tabcolsep{5pt}
\begin{tabular}{@{}lcccc@{}}
\toprule
Baseline & Infeas. & $\sum\Delta T_i$ & Min $V$ & Runtime \\
         & (s)     & (s)              & (kV)    &         \\
\midrule
Uncontrolled    & 57 & --  & 16.35 & --     \\
Static limit    & 0  & 210  & 19.26 & --     \\
Offline GA      & 0  & 7    & 19.01 & 64.9\,s \\
\textbf{Online (this work)} & \textbf{0} & \textbf{7} & \textbf{19.01} & \textbf{1.45\,s} \\
\bottomrule
\end{tabular}
\\[2pt]\footnotesize Peak one-minute RMS is 8.98\,MW for every policy
(Section \ref{sec:discussion}); the online controller advises 1 train (the
high-speed service), one cap level.
\end{table}

\begin{table}[t]
\caption{Estimate against the full solver: section minimum voltage per committed
plan, evaluated on the same second-by-second snapshots. The estimate is optimistic
by only $0.2$\,kV at the binding (uncontrolled) config but by up to $1.5$\,kV once
the trains are lightly capped, which is why the controller screens with a
$1.5$\,kV margin. The three controlled plans each register zero infeasible seconds
because each restores compliance across the whole horizon; only the uncontrolled
case breaches, and estimate and solver agree on that verdict for every plan.}
\label{tab:solvercheck}
\centering\small\setlength\tabcolsep{6pt}
\begin{tabular}{@{}lccc@{}}
\toprule
Policy & Est. min $V$ & Solver min $V$ & Solver infeas. \\
       & (kV)         & (kV)           & (s)            \\
\midrule
Uncontrolled       & 16.57 & 16.35 & 57 \\
Static limit       & 20.71 & 19.26 & 0  \\
Offline GA         & 20.55 & 19.01 & 0  \\
Online controller  & 20.55 & 19.01 & 0  \\
\bottomrule
\end{tabular}
\end{table}

\begin{figure}[t]\centering
\includegraphics[width=\linewidth]{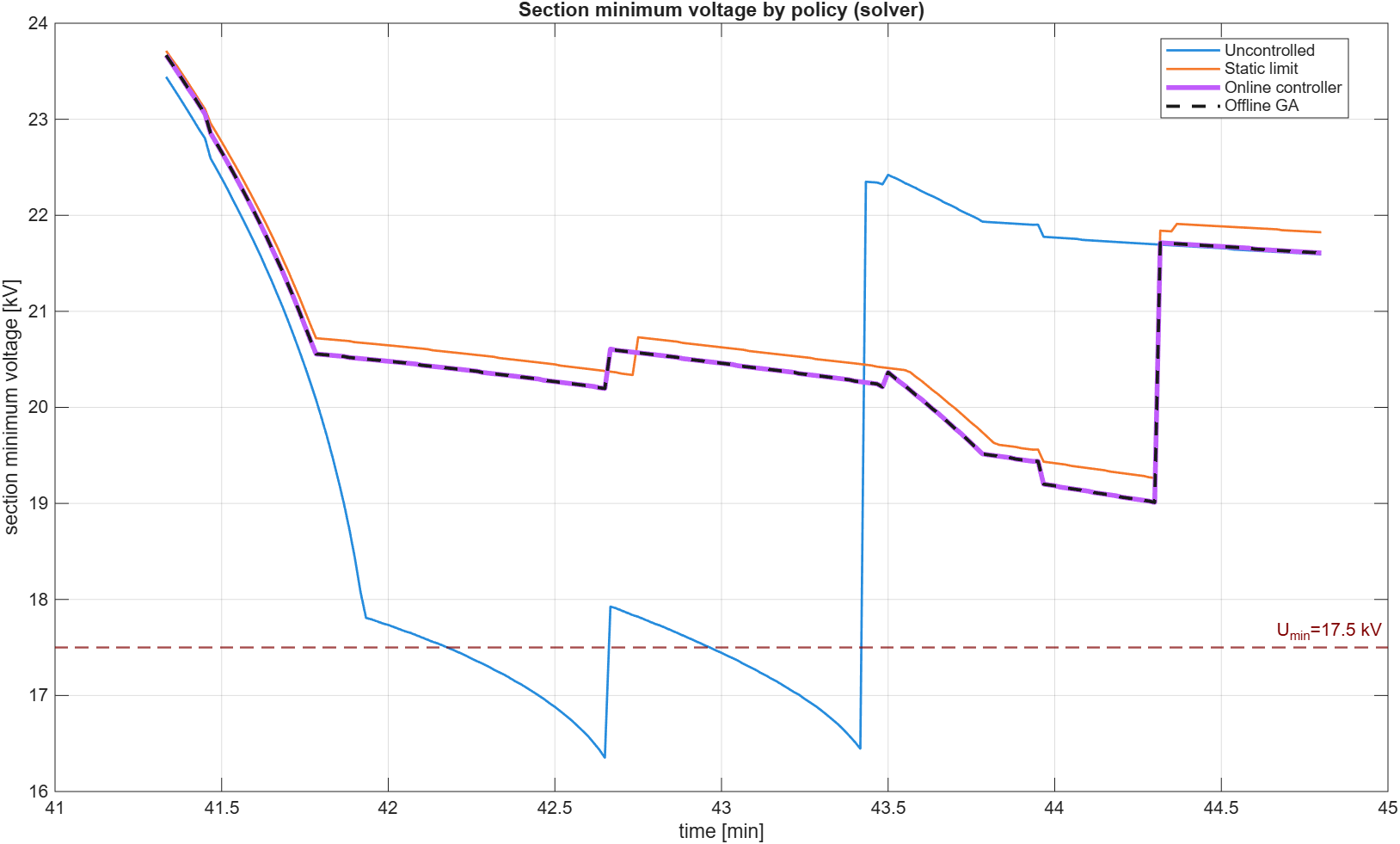}
\caption{Section minimum voltage under each policy, from the full power flow, over
the binding window (the corridor is comfortably above $U_{\min}$ outside it). The
uncontrolled section dips to 16.35\,kV, below the 17.5\,kV compliance line; the
static limit, the GA and the online controller all hold above it, the GA (dashed)
and controller tracing the same trajectory because they commit the same cap. There
is no collapse: feasibility is restored, not a collapse merely softened. This is
the result the paper turns on.}
\label{fig:voltage}
\end{figure}

\begin{figure}[t]\centering
\includegraphics[width=\linewidth]{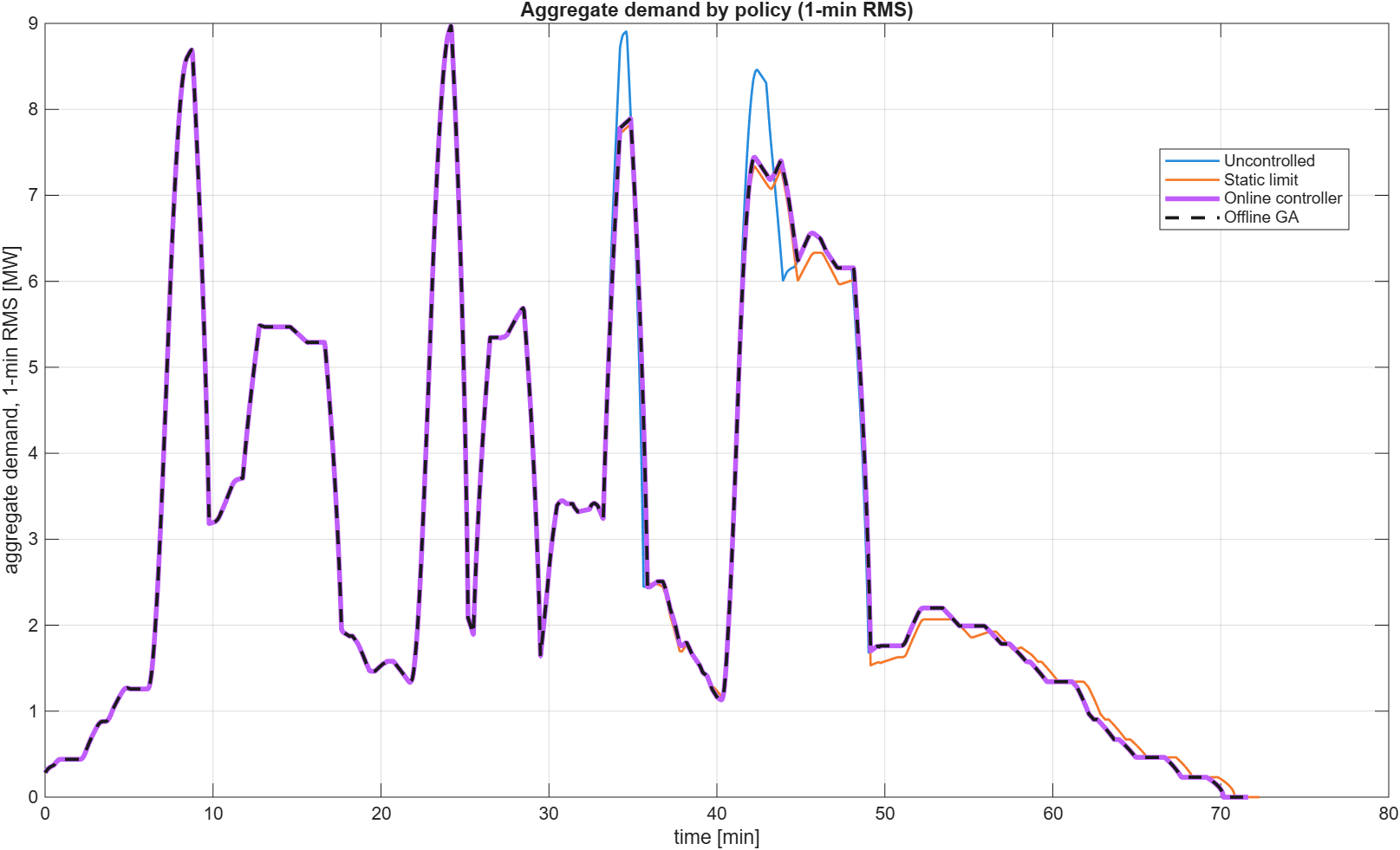}
\caption{Aggregate demand (one-minute RMS) under each policy; the offline GA is
drawn dashed over the online controller, which commits the identical plan. The
peak (8.98\,MW) is the same for every policy: curtailing the far-field binding
train does not move the whole-line aggregate peak, which occurs elsewhere on the
corridor, even as feasibility is restored (Fig. \ref{fig:voltage}). Within the
binding window the static limit reduces demand most, as it caps all six trains
there, whereas the controller and GA reduce it only where the single binding train
is curtailed. This is the secondary ``binding is local'' observation of
Section \ref{sec:discussion} made visible.}
\label{fig:profile}
\end{figure}

\begin{table}[t]
\caption{Hypothesis outcomes, solver-validated. $H_{1a}$: online controller vs
incumbent static limit. $H_{1b}$: online controller vs offline GA optimum.}
\label{tab:hyp}
\centering\small
\begin{tabular}{@{}p{0.17\linewidth}p{0.39\linewidth}p{0.28\linewidth}@{}}
\toprule
Hypothesis & Test & Outcome \\
\midrule
$H_{1a}$ & Restores feasibility at least as well as the static limit, at lower journey cost & Yes: both feasible ($0$ vs $0$\,s), $7$ vs $210$\,s cost, a $30\times$ reduction, one train advised vs the whole section \\[2pt]
$H_{1b}$ & Matches the offline optimum's feasibility and cost, in real time, with few advisories & Yes: identical plan ($0$ vs $0$\,s, $7$ vs $7$\,s cost), $1.45$\,s vs $64.9$\,s runtime, $1$ train advised \\
\bottomrule
\end{tabular}
\end{table}

\noindent\textbf{Reading.} Under outage feeding the uncontrolled section is
infeasible for 57\,s: its far-field voltage falls to 16.35\,kV, below the
17.5\,kV limit, though it does not collapse (Table \ref{tab:headline},
Fig. \ref{fig:voltage}). That is the motivation: normal peak traffic cannot be
supplied within compliance. The incumbent static limit restores feasibility, but
bluntly: capping every train to $0.85$ lifts the minimum to 19.26\,kV at a cost
of 210\,s of aggregate journey time, because it throttles five trains that never
bind alongside the one that does. The offline GA restores the same feasibility
(19.01\,kV) for 7\,s, and it does so by capping the single high-speed service (the
binder) one level. The online controller returns the \emph{identical} plan,
19.01\,kV for 7\,s advising that one train, computed in 1.45\,s against the GA's
64.9\,s. The controller thus dominates the incumbent on cost at equal feasibility
($H_{1a}$, $30\times$) and reaches the offline optimum in real time ($H_{1b}$);
that the GA independently commits the same cap is the strongest available check
that the controller is not cutting a corner. Peak one-minute RMS is $8.98$\,MW for
every policy: curtailing the far-field binder does not move the whole-line
aggregate peak, which sits elsewhere on the corridor, the point developed next.

\section{Discussion}
\label{sec:discussion}

\textbf{The active constraint is local, not aggregate.} The most instructive
observation is that peak one-minute RMS is $8.98$\,MW under every policy while
feasibility varies from 57 to 0 infeasible seconds. Aggregate demand is a
whole-line sum, but the voltage binds locally, in the far field where impedance
from the single-end feed is greatest; curtailing the one train that binds restores
feasibility without materially changing the aggregate. For a power-constrained
outage section, then, the quantity to manage is far-field voltage feasibility, not
gross demand. A blunt demand cap, which acts on the aggregate, is aimed at
the wrong target, which is why it costs so much more journey time than the
location-aware cap for the same relief (210\,s vs 7\,s). \textbf{The estimate is
accurate on average but optimistic at the limit.} Table \ref{tab:solvercheck}
shows the online estimate tracking the solver to within $0.2$\,kV at the binding
config and about $1.5$\,kV once the trains are lightly capped, consistent with
the estimator's reported accuracy \cite{paper1_estimator}, but always on the
optimistic side, and most so at the heavily loaded instants that decide
feasibility. A controller that trusted the estimate at $U_{\min}$ would therefore
under-cap and leave a residual breach. Screening at $U_{\min}+\Delta$ closes this:
with $\Delta=1.5$\,kV the controller's true minimum voltage is 19.01\,kV,
comfortably compliant, at a journey-time cost of only 7\,s. The margin is the
honest, quantified price of a solver-free online screen near the limit, and it is
small. \textbf{Where it helps.} Location-aware curtailment helps most when the
binding is concentrated in identifiable trains, here a single high-speed service;
it approaches the static limit only when loading is uniform, so no small set of
trains dominates the power-by-distance sum. \textbf{Threats to validity.} The
study is in-simulation; per-train power and voltage are inferred from position and
speed rather than measured. The result is a single, solver-selected scenario;
its purpose is to establish that location-aware curtailment restores
\emph{genuine} (solver-confirmed) feasibility more cheaply than a blunt cap, and a
robustness study across traffic levels and binding locations is future work. The
controller is sequential and non-anticipative: it commits each train's cap at
band entry using only information available then and does not revise earlier
commitments, so in denser cases a train already in the band when a later one
binds cannot be re-advised, a residual the offline optimum could in principle
avoid; here the two coincide exactly, so no residual arises, but it bounds the
guarantee in general. \textbf{Operational implication.} Restoring feasibility by
targeted curtailment rather than blanket limiting releases the journey time a
blunt cap wastes, here a factor of thirty, which is what would let more traffic,
including bi-mode services on electric power, run on a constrained section without
a supply upgrade, complementing hardware measures such as wayside energy storage
\cite{ratniyomchai2014recent}.

\section{Conclusion}
\label{sec:conclusion}
In this paper we proposed a near real-time, location-aware controller
that restores the electrical feasibility of a power-constrained AC section by
curtailing only the trains that bind, using the solver-free available-power
estimate of \cite{paper1_estimator} as an in-loop surrogate for the full power
flow, screened with a small voltage margin so that the solver confirms
compliance, and delivered through a cloud-to-edge C-DAS. On a representative GB
25\,kV corridor under outage feeding, selected with the solver in the loop and
evaluated entirely on solver voltages, the incumbent static limit restored
feasibility only by throttling the whole section for 210\,s of journey time,
whereas the controller restored the same feasibility for 7\,s by advising a
single train, a thirtyfold reduction, and reached the offline optimum's exact
solution in 1.45\,s against its minute. A secondary finding is that on such a
section the active constraint is local far-field voltage rather than aggregate
demand, so a blunt demand cap acts on the wrong quantity; a third is that the
online estimate, while accurate on average, is optimistic at the binding instants,
which a fixed screening margin corrects at small cost. The result is established on
a single, solver-selected scenario against real incumbent and theoretical
baselines; a robustness study across traffic levels and binding locations, a
specific-route (e.g.\ East Coast Main Line) deployment study, and a learned
controller for the full receding-horizon problem, for example via multi-agent
reinforcement learning \cite{shang2022energy}, are future work.

\bibliographystyle{IEEEtran}
\bibliography{references}

\end{document}